\documentclass[a4paper]{article}
\usepackage{graphicx}

\newcommand{\ket}[1]{|{#1}\rangle}
\newcommand{\bra}[1]{\langle{#1}|}

\begin{document}

\begin{center}\large
{\bf Performance of quantum phase gates with cold trapped atoms}
    \\[1ex]
A. Negretti$^{1,2,3}$, T. Calarco$^{1,4}$, M. A. Cirone$^{5}$, and A.
Recati$^{2,4}$ 
\\[1ex]
    \normalsize
\em$^1$ECT*, Villa Tambosi Strada delle Tabarelle 286,
I-38050 Villazzano (Trento), Italy
   \\
$^2$Dipartimento di Fisica, Universit\`a di Trento, I-38050,
   Povo, Italy
   \\
$^3$Institut f\"ur Physik, Universit\"at Potsdam, D-14469 Potsdam,
Germany
   \\
$^4$Institut f\"ur Theoretische Physik, Universit\"at Innsbruck,
   A-6020 Innsbruck, Austria
   \\
$^5$Abteilung f\"ur Quantenphysik, Universit\"at Ulm,
   D-89069 Ulm, Germany
\end{center}

\abstract{We examine the performance of a quantum phase gate
implemented with cold neutral atoms in microtraps, when anharmonic
traps are employed and the effects of finite temperature are also
taken into account. Both the anharmonicity and the temperature are
found to pose limitations to the performance of the quantum gate.
We present a quantitative analysis of the problem and show that
the phase gate has a high quality performance for the experimental
values that are presently or in the near future achievable in the
laboratory.}\\

PACS number(s): 03.67.Lx, 32.80.Pj, 34.90.+q

\section{Introduction}

The implementation of quantum logic gates \cite{Nielsen} is a
major goal in the current research in quantum information. Several
schemes have been proposed in the latest years, based on different
physical systems: trapped ions \cite{trio} or neutral atoms
\cite{cala}, cavity--QED and photons \cite{cavi} molecules
\cite{nmr}, quantum dots and Josephson junctions \cite{fort}. The
aim is to implement a fundamental logic quantum gate that works as
a constituent block of a quantum computer \cite{bare}. One of such
gates is the phase gate, whose truth table is

\begin{eqnarray}
\label{tablethrue}
\mid a \rangle \mid a \rangle & \rightarrow & \mid a \rangle \mid a \rangle
\label{trut} \\
\mid a \rangle \mid b \rangle & \rightarrow & \mid a \rangle \mid b \rangle
\nonumber \\
\mid b \rangle \mid a \rangle & \rightarrow & \mid b \rangle \mid a \rangle
\nonumber \\
\mid b \rangle \mid b \rangle & \rightarrow & e^{i\vartheta} \mid b \rangle \mid b \rangle
\nonumber
\end{eqnarray}

Atoms are very good candidates for implementing quantum gates,
because of the significant experimental achievements realized in
recent years. The techniques to cool and trap charged and neutral atoms
have lead to an unprecedented precision in controlling even single
atoms. In particular, neutral atoms seem to be the most promising
systems for quantum information processing, because the dissipative
influence of the environment is relatively weaker when compared to other
physical systems.

A proposal for implementing a phase gate with cold neutral atoms
stored in microtraps has been recently put forward by Calarco {\em et al.}
\cite{cala}. Two atomic internal states, denoted as $\mid \! a \rangle \longmapsto \mid \! 0 \rangle$ and $\mid \! b \rangle \longmapsto \mid \! 1 \rangle$, are used as logical states, and the
operations that realize the truth table Eq.(\ref{trut})
involve the external degrees of freedom. Each atom is placed in a
microtrap, which can be state--selectively switched off and
substituted by a larger harmonic potential that allows collisional
interaction between two atoms. The interaction provides the phase
 that appears in the truth table Eq.(\ref{trut}).
For the sake of simplicity, the traps were assumed to be perfectly harmonic.

In the present article we reexamine this proposal for a phase gate
and in contrast to the work in Ref.\cite{cala}, we employ, when it
is possible and useful, the exact analytic expression of the
eigenstates of two harmonic oscillators with contact interaction
\cite{busc}. Besides, with respect to the harmonic term, we
consider the successive terms of the Taylor series expansion of
the potential. In particular we focus on the fourth term, which
yields the lowest-order correction to the dynamics. The effects of
the temperature are also examined. In realistic experimental
situations these feature will unavoidably become relevant.

The article is organized as follows: In Sec.~\ref{implement} we
describe how the quantum phase gate analyzed here can be
implemented. We stress that two conditions, ({\em i}) full revival
of the vibrational state and ({\em ii}) the acquisition of the
correct phase shift, are the essential ingredients for a correct
performance of the gate. Then section \ref{trapped} describes how
the phase gate can be implemented by using neutral atoms in
microtraps and how the two conditions mentioned above can be
fulfilled. In Sec.~\ref{Ideal} we examine the performance of the
gate when the atoms are at zero temperature and oscillate in
harmonic traps. This situation was already investigated
numerically in \cite{cala}, but here we use the exact eigenstates
of the problem. The results presented here are in total agreement
with those shown in \cite{cala}. In Sec.~\ref{anharmonictraps} we
examine the performance of the gate when the trap is anharmonic.
We obtain quantitative estimates for the quality of the gate
performance. In Sec.~\ref{Tricperformance} we present a heuristic
method that, given a certain anharmonicity, improves the
performance by choosing the trap parameters in such a way as to
optimize the overlap between the initial and the final state. In
Sec.~\ref{temperature} we consider the case when the atoms are at
finite temperature and we provide a quantitative measurement of
the performance using a definition of the fidelity. In
Sec.~\ref{experimental} we show some connections with our
anharmonic model and the physical implementation of the gate on
atom chips. Section \ref{Conclusions} contains our conclusions.

\section{Implementing a phase gate with neutral atoms}
\label{implement} A phase gate with the truth table
Eq.(\ref{trut}) can be implemented by employing two internal
atomic states (hyperfine states) as the logic values $a$ and $b$
and by making use of the vibrational degrees of freedom of the
atoms to manipulate the qubits. In order to keep the exposition
simple, we assume that the atoms are at zero temperature,
described by the state

\begin{eqnarray}
\mid \Psi_{\rm A \rm B}(t=0)\rangle & = &
\mid \psi_{\rm A \rm B}(0) \rangle \otimes \mid \chi \rangle \nonumber \\
& = & \mid \psi_{\rm A \rm B}(0) \rangle
(c_{a}\mid a_{\rm A} \rangle+c_{b}\mid b_{\rm A} \rangle)
(c'_{a}\mid a_{\rm B} \rangle+c'_{b}\mid b_{\rm B} \rangle) \nonumber \\
& = & \mid \psi_{\rm A \rm B}(0) \rangle
(c_{a}c'_{a}\mid a_{\rm A} \rangle \mid a_{\rm B} \rangle
+c_{a}c'_{b}\mid a_{\rm A} \rangle \mid b_{\rm B} \rangle \nonumber \\
& & +c_{b}c'_{a}\mid b_{\rm A} \rangle \mid a_{\rm B} \rangle
+c_{b}c'_{b}\mid b_{\rm A} \rangle \mid b_{\rm B} \rangle),
\label{psi0}
\end{eqnarray}
where

\begin{equation}
\mid \chi \rangle \equiv (c_{a}\mid a_{\rm A} \rangle+c_{b}\mid b_{\rm A} \rangle)
(c'_{a}\mid a_{\rm B} \rangle+c'_{b}\mid b_{\rm B} \rangle)
\label{eqchi}
\end{equation}

is the general initial internal state of the two atoms, the
complex coefficients $c_{a},c_{b},c'_{a},c'_{b}$ satisfy the
normalization conditions $\mid c_{a}\mid ^2+\mid c_{b}\mid ^2=1$
and $\mid c'_{a}\mid ^2+\mid c'_{b}\mid ^2=1$, and $\mid \psi_{\rm
A \rm B}(0) \rangle$ is the vibrational state at $t=0$. The phase
gate operation is obtained with a sequence of unitary
transformations that lead to the final state

\begin{eqnarray}
\mid \Psi_{\rm A \rm B}(t=\tau)\rangle & = &
\mid \psi_{\rm A \rm B}(\tau) \rangle
(c_{a}c'_{a}\mid a_{\rm A} \rangle \mid a_{\rm B} \rangle
+c_{a}c'_{b}\mid a_{\rm A} \rangle \mid b_{\rm B} \rangle \nonumber \\
& & +c_{b}c'_{a}\mid b_{\rm A} \rangle \mid a_{\rm B} \rangle
-c_{b}c'_{b}\mid b_{\rm A} \rangle \mid b_{\rm B} \rangle)
\label{psif}
\end{eqnarray}
at $t=\tau$. A comparison between the two expressions
Eqs.(\ref{psi0}) and (\ref{psif}) for the initial and final states
shows that two ingredients are essential: ({\em i}) a sign change
must occur only in the last term of Eq.(\ref{psif}) and ({\em ii})
the vibrational state must be disentangled from the internal
states at the end of the gate operations. These two conditions are
fulfilled if the motional state, whose wavefunction is
$\psi(x_{\rm A},x_{\rm B},t)$, regains its initial form
$\psi(x_{\rm A},x_{\rm B},0)$ at some later time $\tau$ and
acquires a phase $\pi$ only when both atoms are in the excited
state.

\subsection{Phase gate with two trapped cold atoms}
\label{trapped}

A natural choice to obtain the recurrence
of the initial state $\psi(x_{\rm A},x_{\rm B},0)$
are atoms oscillating in harmonic traps,
where full revivals of wave packets are periodically
observed for any initial state of a single atom.
However, the sign change, i.e., the occurrence of a phase $\pi$
in the wave packet only if both atoms are internally excited, can be
achieved only via an interaction between the atoms that depends
on the internal states. This interaction provokes deviations
from full revivals of the wave packet and works against a correct
performance of the phase gate. The aim of our studies is
to investigate under which conditions
the complete wave packet revival is at least approximately satisfied,
and to evaluate the corresponding gate fidelity.

The system we consider to implement logic quantum gates
is an array of cold bosonic neutral atoms confined in microtraps.
We briefly describe how the gate works at temperature $T=0$.
More details can be found in \cite{cala}.

We assume that bosonic rubidium atoms are employed and use
typical experimental values for the parameters.
For the sake of simplicity, we focus our attention
on only two atoms, under the assumption that we
can restrict our analysis to a one--dimensional system.
For this purpose a strong harmonic confining potential, with frequency $\omega_{\perp}$,
in the transverse directions $y$ and $z$ can be employed.

\begin{figure}[htbp]
\begin{center}
\includegraphics[width=8.0cm]{{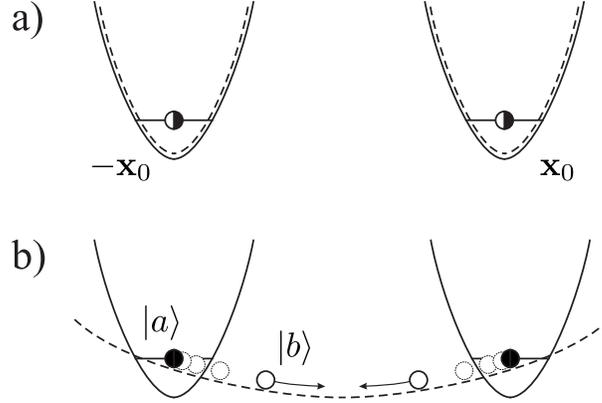}}
\caption{\footnotesize{Configuration at times $t<0$ and $t>\tau$
(a), and during the gate operation (b). The solid (dashed) curves
show the potentials for atoms in internal state $\ket{a}$
($\ket{b}$).}} \label{fig:balls}
\end{center}
\end{figure}

According to Fig.~\ref{fig:balls}, at $t<0$ the two atoms are
confined in two harmonic microtraps of frequency $\omega_{0}$,
centred at $x=-x_{0}$ and $x=x_{0}$, respectively. The distance
between the two traps is such that the atoms do not interact each
other. At time $t=0$, the shape of the trapping potential changes
for the particles in the state $\ket{b}$ into  a common harmonic
well of frequency $\omega<\omega_0$, centred at $x=0$ [dashed line
in Fig.~\ref{fig:balls} (b)], whereas the potentials for the
particles in state $\ket{a}$ remain unchanged [solid line in
Fig.~\ref{fig:balls}(b)]. By removing the barrier, particles in
state $\ket{b}$ start to oscillate and will collide. As a last
step, the atoms have to be restored to the initial motional state
of Fig. \ref{fig:balls}(a). The whole process of switching
potentials is performed through switching the shape of the
potential instantaneously at times $t=0$ and $t=\tau$, where
$\tau$ is a multiple of the oscillation period in the well of
Fig.~\ref{fig:balls}(b) (dashed line).

In order to avoid undesired interactions between the two atoms in
different internal states, the atom in the ground state is shifted
in the transverse direction. Indeed, this interaction would spoil
the performance of the quantum gate, as already discussed in
\cite{cala}. Only when both atoms are excited they oscillate in
the central trap and interact.

For convenience of notation, let us define the following Hamiltonians

\begin{eqnarray}
H_{0}^{-} & = & \frac{p^2}{2M}+\frac{1}{2}M\omega_{0}^2 (x+x_{0})^2
\;\;\;\;\;\;\;\; \label{left} \\
H_{0}^{+} & = & \frac{p^2}{2M}+\frac{1}{2}M\omega_{0}^2 (x-x_{0})^2
\;\;\;\;\;\;\;\; \label{right} \\
H_{0}& = &  \frac{p^2}{2M}+\frac{1}{2}M\omega^2 x^2
\;\;\;\;\;\;\;\; \label{centre} \\
\label{hamiltonians}
H_{\rm N}^{\lambda} & = & H_{0}
+\frac{M^2\omega^3}{\hbar}\lambda x_{\rm N}^4, \;\;\;\;\; {\rm N=A,B}
\;\;\;\;\;\;\;\; \label{oneatom} \\
H_{\rm AB}^{\lambda}& = & H_{\rm A}^{\lambda}+H_{\rm B}^{\lambda}+2 a_{s} \hbar
\omega_{\perp}\delta(x_{\rm A}-x_{\rm B}) \;\;\;\;\;\;\; \label{twoatoms}
\end{eqnarray}
The Hamiltonians (\ref{left}) and (\ref{right}) describe the atoms
oscillating in the harmonic microtraps with frequency
$\omega_{0}$, centred in $x_{0}$ and $-x_{0}$, respectively, where
the two atoms are placed before the gate operation. The atoms
remain in these traps when they are in the internal ground state.
The Hamiltonian (\ref{centre}) describes one atom freely
oscillating in the central trap with frequency $\omega$. The
Hamiltonian (\ref{oneatom}) describes an anharmonic central trap,
derived from the harmonic oscillator Hamiltonian (\ref{centre}) by
adding a quartic term, where the dimensionless parameter $\lambda$
measures the strength of the anharmonicity. The choice of this
particular form of the anharmonic trap will be justified in
Sec.~\ref{anharmonictraps}. Finally, the Hamiltonian
(\ref{twoatoms}) describes two atoms in the anharmonic central
trap interacting via a contact potential described by the Dirac
delta function $\delta(x_{\rm A}-x_{\rm B})$. The coupling
strength depends on the three-dimensional scattering length
$a_{s}$ of the two atoms in the internal $\ket{b}$ and on the
frequency $\omega_{\perp}$ \cite{ols,pertov}. We recall that the
approximation $2 a_{s} \hbar \omega_{\perp}\delta(x_{\rm A}-x_{\rm
B})$ is valid only for
$\l_{\perp}=\left[\hbar/(M\omega_{\perp})\right]^{1/2}\gg a_s$.

The initial state of the two atoms is
\begin{equation}
\mid \psi_{\rm A \rm B}(0) \rangle = \mid \varphi_{0 \rm A}^{-} \rangle
(c_{a}\mid a_{\rm A} \rangle + c_{b}\mid b_{\rm A} \rangle)
\mid \varphi_{0 \rm B}^{+} \rangle
(c'_{a}\mid a_{\rm B} \rangle + c'_{b}\mid b_{\rm B} \rangle)
\end{equation}
where $\varphi_{n}^{\pm}$ denote the eigenstates of $H_{0}^{\pm}$.
The gate operation is described by a unitary evolution operator $U_{\alpha ,\beta}(t)$,
which depends on the internal states $\alpha , \beta=a,b$ of the two
atoms and transforms the initial state into

\begin{eqnarray}
\mid \psi_{\rm A \rm B}(t) \rangle & = & U_{\alpha ,\beta}(t)
\mid \psi_{\rm A \rm B}(0) \rangle \nonumber \\
& = & \left( e^{-i\omega_{0}t} \mid \varphi_{0 \rm A}^{-} \rangle\right)
\left( e^{-i\omega_{0}t} \mid \varphi_{0 \rm B}^{+} \rangle\right)
c_{a}c'_{a}\mid a_{\rm A} \rangle \mid a_{\rm B} \rangle + \nonumber \\
& & \left( e^{-i\omega_{0}t} \mid \varphi_{0\rm A}^{-} \rangle\right)
\left( e^{-\frac{i}{\hbar}H_{\rm B}^{\lambda}t} \mid \varphi_{0 \rm B}^{+} \rangle\right)
c_{a}c'_{b}\mid a_{\rm A} \rangle \mid b_{\rm B} \rangle+ \nonumber \\
& & \left( e^{-\frac{i}{\hbar}H_{\rm A}^{\lambda}t} \mid \varphi_{0 \rm A}^{-} \rangle\right)
\left( e^{-i\omega_{0}t} \mid \varphi_{0\rm B}^{+} \rangle \right)
c_{b}c'_{a}\mid b_{\rm A} \rangle \mid a_{\rm B} \rangle+ \nonumber \\
& & \left( e^{-\frac{i}{\hbar}H_{\rm AB}^{\lambda}t}
\mid \varphi_{0 \rm A}^{-}\rangle \mid \varphi_{0 \rm B}^{+} \rangle \right)
c_{b}c'_{b}\mid b_{\rm A} \rangle \mid b_{\rm B} \rangle
\label{eq.10}
\end{eqnarray}
at a later time $t$. Here the anharmonicity of the central trap has been
taken into account. The state Eq.(\ref{eq.10}) is in general no longer
a separable state of motional and internal degrees of freedom.
However, in the present scheme the separation between
the states of the motional and internal degrees of freedom
can be realized to a good approximation. Indeed, the state

\begin{equation}
\mid \psi_{\rm A\rm B}^{aa}(t) \rangle \equiv
\left( e^{-i\omega_{0}t} \mid \varphi_{0 \rm A}^{-} \rangle\right)
\left( e^{-i\omega_{0}t} \mid \varphi_{0 \rm B}^{+} \rangle\right)
\label{psi00}
\end{equation}
describes two harmonic oscillators in two well-separated
microtraps and has therefore full revivals with period
$T_{osc}^0\equiv 2\pi/\omega_{0}$ for any initial state. The
states

\begin{equation}
\mid \psi_{\rm A \rm B}^{ab}(t) \rangle \equiv
\left( e^{-i\omega_{0}t} \mid \varphi_{0 \rm A}^{-} \rangle\right)
\left( e^{-\frac{i}{\hbar}H_{\rm B}^{\lambda}t} \mid \varphi_{0 \rm B}^{+} \rangle\right)
\label{psi01}
\end{equation}
and

\begin{equation}
\mid \psi_{\rm A \rm B}^{ba}(t) \rangle \equiv
\left( e^{-\frac{i}{\hbar}H_{\rm A}^{\lambda}t} \mid \varphi_{0 \rm A}^{-} \rangle\right)
\left( e^{-i\omega_{0}t} \mid \varphi_{0 \rm B}^{+} \rangle\right)
\label{psi10}
\end{equation}
describe one atom in the microtrap
and the other in the central trap. In this case, the atom
in the microtrap is shifted in the transverse direction
in order to avoid undesired interaction between the atoms.
Therefore, if the central trap is harmonic ($\lambda=0$), full revivals
of the two wave packets at different periods occur.

The state

\begin{equation}
\mid \psi_{\rm A \rm B}^{bb}(t) \rangle \equiv
\left( e^{-\frac{i}{\hbar}H_{\rm AB}^{\lambda}t}
\mid \varphi_{0\rm A}^{-}\rangle \mid \varphi_{0 \rm B}^{+} \rangle \right)
\label{psi11}
\end{equation}
is affected by the interaction between the atoms in the wide trap.
This interaction is necessary in order to yield the sign change
for the phase gate operation but it also modifies the atomic wave
packet. For a good performance of the quantum gate the
modification must be small. The overlap fidelity

\begin{equation}
\label{overlapdef}
O(\psi^{bb}_{\rm A B},t)\equiv \;\;
\langle\psi_{\rm A B}^{bb}(t) \mid \psi_{\rm A B}(0)\rangle
\label{ovfi}
\end{equation}
between the initial vibrational state $\mid \psi_{\rm A B}(0)\rangle$
and that at a later time $t>0$, $\mid \psi_{\rm A B}^{11}(t) \rangle$
gives an estimate of the quality of the gate performance.
If the revival of the motional state
is nearly complete at $t=\tau$, it results

\begin{equation}
\mid O(\psi^{bb}_{\rm A B},\tau) \mid^2 \simeq 1.
\label{over}
\end{equation}
Under this condition, we can write the motional state at time $\tau$ as

\begin{equation}
\mid \psi_{\rm A B}^{bb}(\tau) \rangle \simeq
e^{-i\phi_{bb}(\tau)}\mid \psi_{\rm A B}(0) \rangle
\label{eq11}
\end{equation}
where $\phi_{bb}(\tau)$ denotes the phase of the motional
wave function. When

\begin{equation}
\phi_{bb}(\tau)=\pi,3\pi,5\pi,\ldots,
\label{shif}
\end{equation}
the phase gate operation is correctly realized. In the next
sections we investigate when the two conditions of full or nearly
full revival (expressed by Eq.~(\ref{over})) of the vibrational
state of the excited atoms and the acquistion of the correct phase
(expressed by Eq.~(\ref{shif})) are satisfied.

\section{Ideal case: two cold atoms in harmonic traps}
\label{Ideal}
We examine first the performance of the quantum phase gate when
all traps are harmonic.
This ideal situation was already investigated in \cite{cala},
but we examine it again, because here we use the exact
solutions of the Schr\"odinger equation for this problem.
Indeed, it has been recently found that the problem of two interacting
atoms in one harmonic trap has an exact
solution in one, two and three dimensions \cite{busc}.
Here we simply summarize the results for one dimension.
It is useful to define the new
coordinates $X\equiv (x_{\rm A}+x_{\rm B})/\sqrt{2}$ and
$x\equiv (x_{\rm A}-x_{\rm B})/\sqrt{2}$. The pseudo--particle described
by the centre of mass coordinate $X$ is a free harmonic oscillator,
with eigenstates $\varphi_{n}(X)$ and energy $E_{n}=\hbar \omega
(n+1/2)$. The pseudo--particle described
by the relative coordinate $x$ describes a harmonic oscillator
that feels the contact potential at the origin $x=0$.
Its eigenstates split into two subsets, depending on the
function parity. The odd eigenstates are still those of the free harmonic
oscillator, since the contact potential acts only at $x=0$, where the odd
wave functions vanish. The even solutions

\begin{equation}
\varphi_{\nu}(x)\equiv B_{\nu}\left( \frac{M\omega}{\hbar} \right)^{1/4}
\exp\left[ -\frac{M\omega}{2\hbar}x^2 \right]
U\left( -\nu,\frac{1}{2};\frac{M\omega}{\hbar}x^2 \right)
\label{rela}
\end{equation}
have energy $E_{\nu}\equiv \hbar \omega (2\nu+1/2)$. Here the
real parameters $\nu$ are solution of the transcendental equation

\begin{equation}
\frac{\Gamma(1/2-\nu)}{\Gamma(-\nu)}=-\frac{1}{\sqrt{2}}\frac{a_s}{a_x}
\frac{\omega_{\perp}}{\omega}
\end{equation}
where $a_x \equiv \sqrt{\hbar/(M\omega)}$ is the characteristic
length associated with the central trap and the normalization coefficients

\begin{equation}
B_{\nu}=\sqrt{\frac{\Gamma(1/2-\nu)\Gamma(-\nu)}
{\pi[\psi(1/2-\nu)-\psi(-\nu)]}}
\end{equation}
are defined with the help of the logarithmic derivative $\psi$
of the gamma function.

With the help of these exact eigenstates we evaluate the fidelity
Eq.~(\ref{over}) and the phase shift Eq.~(\ref{shif}) when both
atoms are excited. At $t<0$ the two atoms are in the vibrational
ground states of their own microtrap, say, atom A in the left
trap, centred at $x=-x_{0}$, and atom B in the right trap, centred
at $x=x_{0}$. At $t=0$  the microtraps are switched off, the
central trap is switched on and both atoms oscillate in the same
trap and interact. In order to calculate the overlap fidelity
Eq.~(\ref{over}), we can proceed in two different but equivalent
ways. Since the atoms are identical, the gate performs correctly
even when the two atoms end up in the other trap at the end of the
operation. We can therefore either symmetrize the initial wave
function for the two bosonic atoms, or estimate the overlap
fidelity Eq.~(\ref{over}) as the sum of the probabilities to find
the (distinguishable) atoms in the original traps or with the
initial positions interchanged. Since in this section we use the
wave functions of the center of mass $X$ and of the relative
coordinate $x$, the first approach is more convenient. The
two--atom vibrational state at generic time $t>0$ is

\begin{equation}
\psi(X,x,t)=\sum_{n,\nu}c_{n,\nu}e^{-i(n+2\nu+1)\omega t}\varphi_{n}(X)
\varphi_{\nu}(x)
\end{equation}
where the odd eigenstates of the relative coordinate $x$ are not included
since they describe fermions. The coefficients $c_{n,\nu}$ vanish when $n$ is
odd, otherwise

\begin{equation}
c_{n,\nu}=
2\pi^{-1/4}\sqrt{\frac{\omega_0}{n(\omega_0+\omega)}}
e^{-\frac{M\omega_0x_0^2}{\hbar}}2^{-n/2}\frac{\sqrt{n!}}{(n/2)!}
\left(\frac{\omega-\omega_0}{\omega+\omega_0}\right)^{n/2}B_{\nu}
{\mathcal I}_{\nu}
\end{equation}
where

\begin{equation}
{\mathcal I}_{\nu}=\int_{-\infty}^{\infty}dy
e^{-\left(1+\frac{\omega_0}{\omega}\right)\frac{y^2}{2}}
U\left(-\nu,1/2;y^2\right)
\cosh\left[x_0\omega_0\sqrt{\frac{2M}{\hbar\omega}}y\right]
\end{equation}

\begin{figure}[htbp]
\begin{center}
\includegraphics[width=10.0cm,height=6.7cm]{{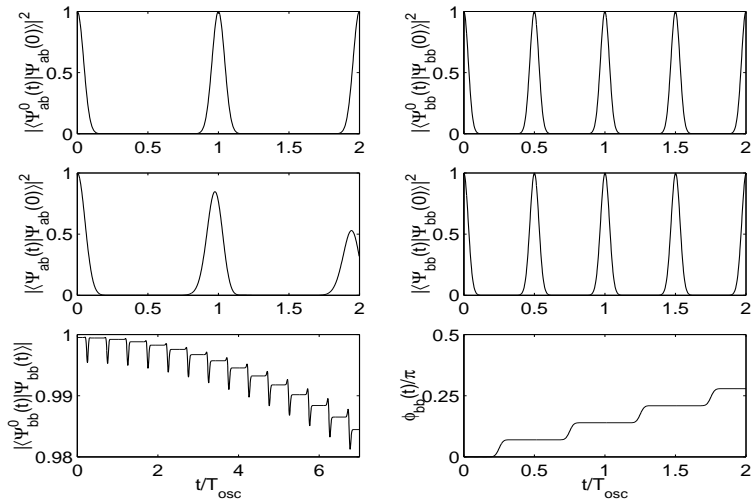}}
\caption{\footnotesize{Dynamics during gate operation: projection
of the initial state on the state evolved without (top) and with
interaction (center); projection of the evolved state on the
corresponding state evolved without interaction (bottom left) and
interaction-induced phase shift (bottom right). We choose
$\omega=2\pi17.23$kHz and $\omega_{\perp}=2\pi150$kHz with the
initial wells having a frequency $\omega_0=2\omega$ and displaced
by $x_0=410$ nm.} This reproduces analytically the results
obtained in \cite{cala} with a purely numerical approach.}
\label{fig:idealgate}
\end{center}
\end{figure}

In Fig.~\ref{fig:idealgate} we show the overlap fidelity,
Eq.~(\ref{over}). The revival of the vibrational state, occurs
with periodicity $T_{osc}/2$. Indeed, after this period the two
atoms are in $x=\pm x_{0}$. The revival is nearly complete, as the
overlap fidelity approaches the value 0.99. In
Fig.~\ref{fig:idealgate} we also report the situation in which
atoms in different internal states feel a contact potential in
order to show what happens.

In Fig.~\ref{fig:idealgate} we show the phase shift
$\phi_{bb}(\tau)$ due to the interaction. There is a fast change
of the phase between the two revivals, in correspondence to the
presence of the two atoms at the bottom of the trap, where the
interaction occurs. The figure suggests that one can assume that
at each interaction there is a jump in the phase. Therefore, after
a suitable number of collisions, the vibrational state acquires
the correct phase for the gate operation. The exact results shown
here confirm the validity and accuracy of the results of
Ref.\cite{cala} obtained numerically.

\section{Phase gate performance with anharmonic traps}
\label{anharmonictraps} The operation of the phase gate relies on
several simplifying assumptions. In the previous section we have
assumed that the atoms are at zero temperature and oscillate in
harmonic traps. The experimental conditions are necessarily less
ideal and the problem of estimating the effect of deviations from
the ideal conditions is particularly important. Among the causes
that can lead to bad performance of the phase gate, we mention
random noise, caused by fluctuating electromagnetic fields,
temperature effects and anharmonicity of the trapping potentials.
In particular, some anharmonicity might more easily appear in the
wider central trap and seems to be the most important disturbance
to be taken into account.

In this section we study the gate performance when the excited
atoms oscillate in an anharmonic trap, as described by the
Hamiltonians Eqs.~(\ref{oneatom}) and (\ref{twoatoms}). Indeed,
independently of its exact expression, we can expand the trapping
potential in Taylor series. The first correction to the harmonic
approximation is a cubic term, that we shall neglect since at
first order of approximation it does not lead to any correction to
the energy and it does not affect the atomic motion.

The next relevant correction to the harmonic trap is
a quartic term. It is symmetric and it is responsible for an
energy shift. We neglect the other terms in the Taylor expansion,
which yield minor perturbations. The atom dynamics in the
central trap is then described by the Hamiltonians (\ref{oneatom})
and (\ref{twoatoms}).

The atomic wave packet does no longer show full revivals.
Moreover, the partial revivals of the initial state that still
occur are no longer periodic, in a strict sense. Therefore, the
question arises as to how much anharmonicity can be tolerated
without destroying the good performance of the phase gate.

\subsection{Anharmonic trap: Overlap fidelity for the internal states
$\mid a_{\rm A} \rangle \mid b_{\rm B} \rangle$ and $\mid b_{\rm A} \rangle
\mid a_{\rm B} \rangle$}

In this subsection we examine the performance of the gate when
only one atom is in the internal excited state. During the
gate operation the excited atom
(atom A in the left trap, say) oscillates freely in the
central anharmonic trap, while atom B oscillates as a
free harmonic oscillator in its microtrap.
We need only to focus our attention on the motion of atom A,
whose vibrational state at time $t$ can be expanded on eigenstates of the
harmonic oscillators according to

\begin{equation}
\mid \psi_{\rm A}(t) \rangle=\sum_{n}c_{n}(t)e^{-i2\pi nt} \mid
\varphi_{n \rm A} \rangle\qquad \left(t\longrightarrow
\frac{t}{T_{osc}}\right)
\end{equation}
where $\varphi_{n}$ denotes the eigenstates of $H_{0}$ and
$T_{osc}$ is the period of oscillation of the central trap [dashed
line in Fig.~\ref{fig:balls}(b)]. The expansion coefficients
$c_{k}(t)$ satisfy the differential equations

\begin{eqnarray}
\label{coef01}
\dot{c}_{n}(t) & = -i\frac{\pi}{2}\lambda & \left\{\sqrt{(n+1)(n+2)(n+3)(n+4)}c_{n+4}(t)e^{-i8\pi t}\right.\nonumber\\& & +2(2n+3)\sqrt{(n+1)(n+2)}c_{n+2}(t)e^{-i4\pi t}\nonumber\\& & +3\left[(n+1)^2+n^2\right]c_{n}(t)\nonumber\\& & +2(2n-1)\sqrt{(n-1)n}c_{n-2}(t)e^{i4\pi t}\nonumber\\& & +\left.\sqrt{(n-3)(n-2)(n-1)n}c_{n-4}(t)e^{i8\pi t}\right\},
\end{eqnarray}

\begin{figure}[htbp]
\begin{center}
\includegraphics[width=8.0cm,height=5.0cm]{{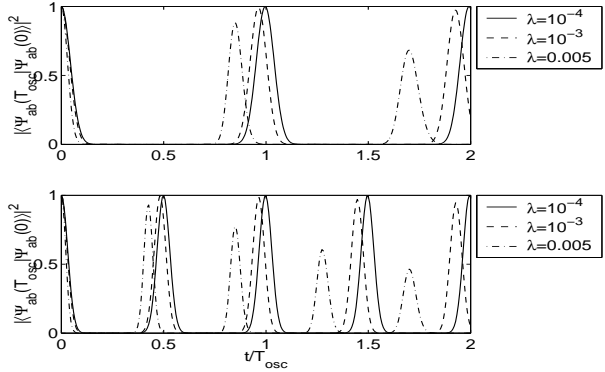}}
\caption{\footnotesize{Dynamics during gate operation: projection
of the initial state on the state evolved with different values of
anharmonicity. Trap parameters have the same values as in
Fig.~\ref{fig:idealgate}.}} \label{fig:anharmonicgate}
\end{center}
\end{figure}

The overlap fidelity is

\begin{equation}
\mid O^{ba}(t) \mid^2 \equiv \mid \langle \varphi_{0 \rm A}^{-} \mid
e^{-\frac{i}{\hbar}H_{\rm A}^{\lambda}t} \mid \varphi_{0 \rm A}^{-} \rangle \mid^2
=\mid \sum_{k}c_{k}(t)e^{-i2\pi kt} c_{k}^{*}(0) \mid^2
\end{equation}
where

\begin{equation}
c_{k}(0) \equiv \; \langle \varphi_{k \rm A} \mid \varphi_{0 \rm A}^{-} \rangle
\end{equation}

The overlap fidelity (\ref{ovfi}) has been evaluated numerically
and is plotted in Fig.~\ref{fig:anharmonicgate} on the top for
different values of the parameter $\lambda$. It is evident from
the figure that the gate tolerates some anharmonicity, with a
threshold value of the order of $\lambda \sim 10^{-4}$. The
specular case of atom B excited, while atom A is not excited,
gives the same results.

\subsection{Anharmonic trap: Overlap fidelity for the internal state
$\mid b_{\rm A} \rangle \mid b_{\rm B} \rangle$}

Here we examine the performance of the gate when both atoms are
excited. We expand the motional state of the two atoms on
the eigenstates of the central harmonic trap

\begin{equation}
\mid \psi_{\rm A B}(t) \rangle=\sum_{k,l}c_{k,l}(t)e^{-i2\pi
(k+l)t} \mid \varphi_{k \rm A} \rangle \mid \varphi_{l \rm B}
\rangle\qquad \left(t\longrightarrow \frac{t}{T_{osc}}\right)
\end{equation}
and the expansion coefficients $c_{k,l}(t)$ statisfy the equations

\begin{eqnarray}
\dot{c}_{k,l}(t) & = & -i4\frac{\omega_{\perp}}{\omega}a_s\sqrt{\frac{m\omega}{\hbar}}
e^{i(k+l)2\pi t} \sum_{n,m}c_{n,m}(t)e^{-i(n+m)2\pi t}A_{klnm} \nonumber \\
& & -i\frac{\pi}{2}\lambda \left[ \sqrt{k(k-1)(k-2)(k-3)} \right.
c_{k-4,l}(t)e^{8i\pi t} \nonumber \\
& & +(4k-2)\sqrt{k(k-1)}c_{k-2,l}(t)e^{4i\pi t}
+(6k^2+6k+3)c_{k,l}(t) \nonumber \\
& & +(4k+6)\sqrt{(k+1)(k+2)}c_{k+2,l}(t)e^{-4i\pi t}
\nonumber \\
& & + \sqrt{(k+1)(k+2)(k+3)(k+4)}c_{k+4,l}(t)e^{-8i\pi t} \nonumber \\
& & + \sqrt{l(l-1)(l-2)(l-3)}c_{k,l-4}(t)e^{8i\pi t} \nonumber \\
& & +(4l-2)\sqrt{l(l-1)}c_{k,l-2}(t)e^{4i\pi t}
+(6l^2+6l+3)c_{k,l}(t)\nonumber \\
& & +(4l+6)\sqrt{(l+1)(l+2)}c_{k,l+2}(t)e^{-4i\pi t}
\nonumber \\
& & \left. + \sqrt{(l+1)(l+2)(l+3)(l+4)}c_{k,l+4}(t)e^{-8i\pi t} \right]
\label{ckl}
\end{eqnarray}
where

\begin{equation}
A_{klnm}\equiv \frac{1}{\sqrt{2^{k+l+n+m}}\sqrt{k!l!n!m!}}\int_{-\infty}^{\infty}
d\xi e^{-2\xi^2}H_{k}(\xi)H_{l}(\xi)H_{n}(\xi)H_{m}(\xi)
\end{equation}
and $H_{n}$ denotes the Hermite polynomial of order $n$. The first
term in Eq.~(\ref{ckl}) comes from the contact interaction between
the two atoms, whereas the other terms are due to anharmonicity.
Assuming that the atoms are distinguishable, the overlap fidelity
is

\begin{eqnarray}
O^{bb}(t) & = & \mid \langle \varphi_{0A}^{-} \varphi_{0B}^{+} \mid
e^{-\frac{i}{\hbar}H_{AB}^{\lambda}t} \mid \varphi_{0A}^{-}\varphi_{0B}^{+} \rangle \mid^2
\nonumber \\
& & + \mid \langle \varphi_{0A}^{+}\varphi_{0B}^{-} \mid
e^{-\frac{i}{\hbar}H_{AB}^{\lambda}t} \mid \varphi_{0A}^{-}\varphi_{0B}^{+} \rangle \mid^2
\nonumber \\
& = & \mid \sum_{kl}c_{kl}(t)e^{-ik2\pi t}e^{-il2\pi t}
c_{kl}^{AB*}(0) \mid^2\nonumber \\
& & +\mid \sum_{kl}c_{kl}(t)e^{-ik2\pi t}e^{-il2\pi t}
c_{kl}^{BA*}(0) \mid^2
\end{eqnarray}
where we have defined

\begin{eqnarray}
c_{kl}^{AB*}(0) & \equiv & \langle \varphi_{0A}^{-}\varphi_{0B}^{+} \mid
\varphi_{kA}\varphi_{lB} \rangle \nonumber \\
c_{kl}^{BA*}(0) & \equiv & \langle \varphi_{0A}^{+}\varphi_{0B}^{-} \mid
\varphi_{kA}\varphi_{lB} \rangle
\end{eqnarray}
We have numerically evaluated the overlap fidelity, which is shown
in Fig.~\ref{fig:anharmonicgate} on the bottom for different
values of the parameter $\lambda$. Also in this case we see that
an anharmonicity of the order of $\lambda \sim 10^{-4}$ or less
does not prejudicate the performance of the phase gate.

In conclusion of this section, we note that different choices of
the values of the parameters lead to very different performance
qualities. For a fixed value of the anharmonicity parameter
$\lambda$ different gate performances are obtained for different
values of the other parameters. From the first term on the right
hand side of Eq.~(\ref{ckl}) one sees that the effect of the
contact interaction on the atom dynamics depends on the value of
$\omega_{\perp}/\omega a_s\left(M\omega/\hbar\right)^{1/2}$. If
this quantity is larger than $\approx 0.7$, it spoils the
occurrence of full revivals; if it is too small, too many
oscillations are needed to create the phase $\phi_{bb}(t)=\pi$. We
also note that the frequency $\omega_{\perp}$ must be large enough
to prevent excitations along the transverse direction
($\l_{\perp}\gg a_s$). In spite of these limitations, it is
possible to find reasonable values for these parameters that make
a correct performance possible, as the data in
Fig.~\ref{fig:anharmonicgate} (bottom) show.

\section{Optimization of gate performance}
\label{Tricperformance}
\begin{figure}[htbp]
\begin{center}
\includegraphics[width=8.0cm,height=5.0cm]{{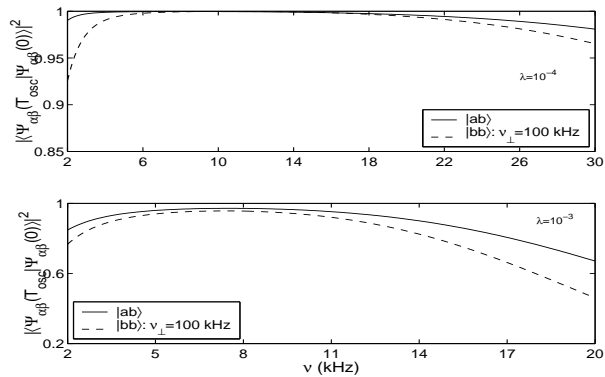}}
\caption{\footnotesize{Overlaps
$\left|O\left(\psi_{\alpha\beta},\tau\right)\right|$ in order to
optimize the performance of the phase gate.}} \label{fig:trick}
\end{center}
\end{figure}

The gate perfomance can be optimized reducing the number of oscillations and selecting the trap frequencies $\omega(\lambda)$ and $\omega_{\perp}(\lambda)$ such that the overlaps $\left|O\left(\psi_{\alpha\beta},\tau\right)\right|$ are close to one. In this way we have two effects: better performance and faster gate. In Fig. \ref{fig:trick} we report the overlaps $\left|O\left(\psi_{\alpha\beta},\tau\right)\right|$ for two different anharmonic situations.

In Table \ref{tab:frequencies} we report the values that optimize the performance.

\begin{table}[htbp]
\begin{center}
\begin{tabular}{||*{4}{c||}}
\hline
$\lambda$ & $\omega$ (kHz) & $\omega_0$ (kHz) & $\omega_{\perp}$ (kHz) \\
\hline
\hline
$10^{-4}$ & $2\pi\cdot 12.00$ & $4\cdot\omega$ & $2\pi\cdot 769.00$ \\
\hline
$10^{-3}$ & $2\pi\cdot 4.00$ & $4\cdot\omega$ & $2\pi\cdot 297.70$ \\
\hline
\end{tabular}
\end{center}
\caption{\footnotesize{Trap frequencies maximizing the fidelity
curves in Fig.~\ref{fig:trick}.}} \label{tab:frequencies}
\end{table}

\begin{figure}[htbp]
\begin{center}
\includegraphics[width=8.0cm,height=5.5cm]{{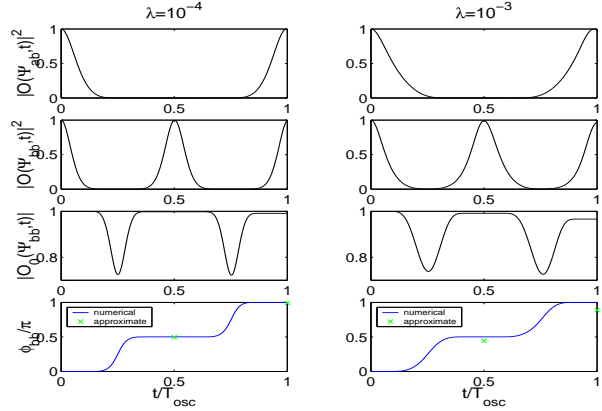}}
\caption{\footnotesize{Dynamics during gate operation: projection
of the initial state on the state evolved with interaction (first
two rows); projection of the evolved state on the corresponding
state evolved without interaction (third row) and
interaction-induced phase shift (bottom). The displacement of the
two initial separated wells is the same as in
Fig.~\ref{fig:idealgate}.}} \label{fig:gateanharmonic}
\end{center}
\end{figure}

With these values we obtain the results shown in Fig.
\ref{fig:gateanharmonic}. The crosses in the bottom pictures of
Fig. \ref{fig:gateanharmonic} are obtained by means of these two
assumptions: (i) the particles move against each other, come in
contact during a certain time interval $\left[t_i,t_F\right]$ and
then separate again; and (ii) the velocity of each particle and
the shape of its wave function do not vary during the interaction.
It follows that (see Ref.\cite{cala} for more details)

\begin{equation}
\label{phivel}
\phi_{bb}\left(T_{osc}\right)=2\frac{\omega_{\perp}a_{s}}{\omega v}
\end{equation}
in harmonic oscillator units. Here the velocity $v$ is a positive constant value given by
\begin{eqnarray}
v&=&\left|\partial_t\bra{\psi_{\pm}(t)}x\ket{\psi_{\pm}(t)}|_{t=t_k}\right|=\nonumber\\
&=&2\mathcal{R}\left\{\sum_n\overline{c_n\left(t_k\right)}E^0_n\left[\sqrt{\frac{n+1}{2}}c_{n+1}\left(t_k\right)-\sqrt{\frac{n}{2}}c_{n-1}\left(t_k\right)\right]+\right.\nonumber\\
&+&\left. i\lambda\sum_{n,q}\overline{c_n\left(t_k\right)}c_q\left(t_k\right)e^{i(n-q)\frac{\pi}{2}}\left[\sqrt{\frac{n+1}{2}}\mathcal{I}_{n+1,q}^4+\sqrt{\frac{n}{2}}\mathcal{I}_{n-1,q}^4\right]\right\}\nonumber,
\end{eqnarray}
where $\mathcal{I}_{n+1,q}^4$ and $\mathcal{I}_{n-1,q}^4$ are given by
\begin{eqnarray}
\mathcal{I}_{n,q}^s=\left[2^{n+q}n!q!\pi\right]^{-1/2}\int_{-\infty}^{+\infty}dx e^{-x^2}x^sH_n(x)H_q(x),
\end{eqnarray}
whereas the coefficients $c_n(t)$ are given by equation
(\ref{coef01}) and $t_k=\left(2k+1\right)T_{osc}/4$ with $k$ an
integer.

Fig.~\ref{fig:gateanharmonic} shows that there is a good agreement
between the numerical result (solid line) and that given by
equation (\ref{phivel}). The agreement is not perfect, though,
simply because the velocity is not constant during the
interaction.

\section{Gate performance in anharmonic traps at finite temperature}
\label{temperature} Now we examine the gate performance when both
anharmonicity and temperature effects are taken into account.
Since the atoms are not in the ground state of the initial
trapping potential and therefore their temperature $T\neq 0$, it
follows that, for a finite temperature $T$, the initial state of
the two atoms is described by the density matrix

\begin{equation}
\rho_{0} = \frac{1}{Z} \sum_{k,l=0}^{\infty}
P_{k,l}(T)
\mid \varphi_{k \rm A}^{-} \rangle \mid \varphi_{l \rm B}^{+} \rangle \otimes
\langle \varphi_{k \rm A}^{-} \mid \langle \varphi_{l \rm B}^{+} \mid
\label{dens}
\end{equation}
where the occupation probabilities of the $k$ and $l$ states are
calculated assuming, for each atom, a thermal distribution
corresponding to temperature $T$, as expressed by

\begin{equation}
\label{prob}
P_{k,l}(T)\equiv \exp \left[ -\frac{\hbar \omega_{0}}{k_{B}T}(k+l) \right]
\end{equation}
and

\begin{equation}
Z=\sum_{k,l=0}^{\infty}P_{k,l}(T)
\end{equation}
is the canonical partition function.

\subsection{Gate fidelity}

The most general logical input state has the form
\begin{equation}
\ket{\chi}=\sum_{\alpha,\beta=0}^1c_{\alpha\beta}\ket{\alpha,\beta},
\end{equation}
which is an arbitrary superposition of all two-qubit computational basis states. The goal of gate operation is to obtain the ideal output
\begin{equation}
\ket{\tilde{\chi}}=\sum_{\alpha,\beta=0}^1c_{\alpha\beta}e^{i\phi_{\alpha\beta}}\ket{\alpha,\beta}.
\end{equation}
This is equivalent to the desired two-qubit transformation Eq.
(\ref{tablethrue}): provided that
$\vartheta=\phi_{00}+\phi_{11}-\phi_{01}-\phi_{10}$, the one can be
recovered from the other by redefining the logical states via
qubit rotations.

Since in this case the two atoms are described by a density
matrix, we cannot use the overlap fidelity condition
Eq.~(\ref{over}) to estimate the performance of the phase gate. We
use therefore the minimum fidelity \cite{schumi} to characterize
the quality of the phase gate performance \cite{cala},
\begin{eqnarray}
F & = & \min_{\chi} F(\chi) \nonumber \\
& = & \min_{\chi} \langle \tilde{\chi} \mid
{\rm Tr}_{\rm ext}[\mathcal{U}S (\rho_{\rm int} \otimes \rho_{0}) S^{\dagger}
\mathcal{U}^{\dagger}] \mid \tilde{\chi} \rangle
\label{minf}
\end{eqnarray}
Here $\rho_{\rm int}=\mid \chi \rangle \langle \chi \mid$ is the
density matrix of the internal state Eq.~(\ref{eqchi}) and $S$
denotes an operator that simmetrizes the atomic state. If we write
\begin{equation}
\ket{\chi}=\sum_{n=0}^3c_n\ket{n}.
\end{equation}
and assume
\begin{equation}
\mathcal{U}\left[\ket{n}\otimes\rho\right]\approx \ket{n}\otimes\mathcal{U}\left[\rho\right],
\end{equation}
the fidelity takes the form
\begin{equation}
\label{traxfid}
F=\min_{\left\{c_{n}\right\}}\sum_{n,k}\left|c_n\right|^2\left|c_k\right|^2T_{nk},
\end{equation}
where
\begin{equation}
\label{matriceT}
T_{nk}=e^{i\left(\phi_{n}-\phi_{k}\right)}\sum_{n_1,n_2}P_{n_1n_2}(T)\langle n_2,n_1|\mathcal{U}_{n}^\dag\mathcal{U}_{k}|n_1,n_2\rangle.
\end{equation}
The minimum of the fidelity $F(\chi)$ is evaluated in the
Appendix. For the ideal case, that is, without anharmonicity but
with the exact solutions given by (\ref{rela}), the fidelity at
$T=0$ is $F\approx 0.99$.

In Table \ref{tab:fid0K} we show the values of the fidelity at
zero Kelvin with the frequencies given in the Table
\ref{tab:frequencies}, which optimizes the fidelity, for different
values of the anharmonicity. It is important to note that these
results are obtained considering $\tau=T_{osc}$ instead of
$\tau=7T_{osc}$ as in Ref.\cite{cala}, in order to improve the
gate time operation.

\begin{figure}[htbp]
\begin{center}
\includegraphics[width=7.0cm,height=4.5cm]{{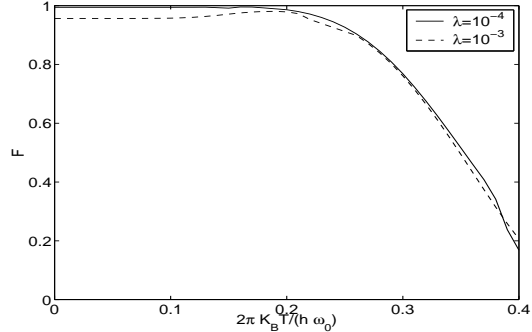}}
\caption{\footnotesize{Fidelity $F$ against temperature
$k_BT/(\hbar\omega_0(\lambda))$ for $^{87}$Rb. Trap parameters
have the values given in Table \ref{tab:frequencies}.}}
\label{fig:fidelityanah}
\end{center}
\end{figure}

\begin{table}[htbp]
\begin{center}
\begin{tabular}{||*{2}{c||}}
\hline
$\lambda$ & $F$\\
\hline
\hline
$10^{-4}$ & $0.99$ \\
\hline
$10^{-3}$ & $0.96$ \\
\hline
\end{tabular}
\end{center}
\caption{\footnotesize{Fidelity at $T=0$ for two values of anharmonicity.}}
\label{tab:fid0K}
\end{table}
When the atoms are at finite temperature, the fidelity decreases.
If we define
$\gamma=\exp\left[-\hbar\omega_0/\left(k_BT\right)\right]$ and
evalute (\ref{matriceT}) neglecting terms of the order
$O\left(\gamma^3\right)$ the fidelity turns out to be $F\approx
0.97$ for $\lambda=10^{-4}$ at $T\approx 0.5\mu$K. In
Fig.~\ref{fig:fidelityanah} we show the behavior of the fidelity
with the temperature for $\lambda=10^{-4}$ and $\lambda=10^{-3}$.

It is important to highlight that, since different values of
$\lambda$ lead to different optimal trap frequencies
$\omega_0(\lambda)$, the curves in the Fig.~\ref{fig:fidelityanah}
are plotted as a function of the ratio
$k_BT/(\hbar\omega_0(\lambda))$. Thus, e.g., the maximum value of
the $x$ axis of the $\lambda=10^{-4}$ curve corresponds at a
temperature of the order of about 1 $\mu$K and for
$\lambda=10^{-3}$ is 0.3 $\mu$K, almost an order of magnitude
smaller.

\section{Estimating $\lambda$ in a realistic situation}
\label{experimental} We show how $\lambda$ is related to the
trap's parameters and the properties of the atoms used in actual
experiments where magnetic traps are used for the confinement of
neutral atoms. The interaction between the magnetic dipole moment
of an atom in some hyperfine state $\ket{F,m_F}$ and an external
magnetic field $\bf{B}$ is
\begin{equation}
H_{{\rm{int}}}=-\bf{\mu}\cdot\bf{B}.
\end{equation}
In an inhomogeneous magnetic field, if the atomic motion is slow
as compared with the velocity of change of the field vector as
seen by the moving atom, the interaction only depends on the
absolute value of the field:
\begin{equation}
H_{{\rm{int}}}=-\mu_zB=g_Fm_F\mu_BB,
\end{equation}
where $\mu_B$ is the Bohr magneton and $g_F$ is the Land\'e
factor.
As in \cite{cala}, we consider here an atomic mirror like the one
realized \cite{mirror,mirror2} from a solid-state magnetic medium
with permanent sinusoidal magnetization
${\bf{M}}=\left(M_0\cos\left[k_Mx,0,0\right]\right)$ along the $x$
axis. In order to avoid trap losses, due to spin flips occuring at
magnetic field zeros \cite{mirror3}, it is necessary to apply a
certain external bias field $B_2$ along the $y$ direction.
Moreover, to obtain a corrugation in the magnetic field modulus,
we add a rotating external field $B_1$ in the $xz$ plane, at an
angle $\theta$ with the surface that can be varied at will. In
this case the magnetic trapping potential is
\begin{eqnarray}
\label{potmagn} V_{m_F}(\mathbf{x})&=&g_F\mu_Bm_F\left\{
\left[B_0e^{-zk_M}\cos^2\left(k_Mx\right)+B_1\cos\theta\right]^2+
\right.\nonumber \\
&&\left.\left[B_0e^{-zk_M}\sin^2\left(k_Mx\right)+B_1\sin\theta\right]^2+B_2^2\right\}^{1/2}.
\end{eqnarray}
where $B_0=\mu_0M_0\left(1-e^{-k_M\delta}\right)/2$ and $\delta$
is the tape thickness. The minima of $V_{m_F}(\mathbf{x})$ form a
periodic pattern above the tape surface, at a height
$z_0=k_M\ln(B_0/B_1)$. The spacing between two nearest minima
along $x$ is just the period of the magnetization $\delta
x_M\equiv 2\pi/k_M$. From the series expansion of
Eq.~(\ref{potmagn}) along $x$ for $\theta=0$ around its minimum,
we obtain for the anharmonicity parameter
\begin{equation}
\lambda=\frac{\pi^3\hbar k_M}{\sqrt{m\mu_B
B_2}}\left(\frac{B_1}{B_2}+\frac{B_2}{3B_1}\right)
\end{equation}
which, for $2\pi k_M^{-1}$ around a few $\mu$m and for fields
$B_i$ of a couple hundred Gauss, is of the order of $10^{-3}$.

\section{Conclusions}
\label{Conclusions}
In the present paper we have extended the investigations concerning
the performance of a phase gate, as proposed by Calarco {\em et al.}
in Ref.\cite{cala}. The phase gate employs cold trapped neutral
atoms and the gate operation is obtained with internal state--selective
trapping potentials that allow collisional interaction between the atoms.

A correct performance of quantum gates is an essential ingredient
of a quantum computer. We have therefore relaxed the ideal
conditions in Ref.\cite{cala} in order to check the tolerance of
the proposed scheme to experimental imperfections. We have
considered the effects of two possible sources of undesired
disturbance, i.e., non-perfectly harmonic trapping potentials and
temperature. The most crucial parameter is the trap anharmonicity
$\lambda$. By studying the dependence of the gate fidelity on such
parameter, we have been able to give a prescription to adjust
other trap parameters as to compensate for this source of
infidelity. However, we found a critical value for $\lambda$
around $10^{-3}$ where the fidelity starts to be significantly
degraded (i.e., well beyond any threshold for fault-tolerant
quantum computation). This value turns out to be right on the edge
of what can be presently achieved with magnetic micropotentials
(atom chips). Thus a further optimization is needed.

The gate performance could be improved by changing the shape of
the trapping potential during the gate operation. This requires
the simultaneous variation of various physical parameters,
$\lambda$, $\omega$, $\omega_0$, and $\omega_{\perp}$, since the
fidelity depends on all these quantities. A viable approach to
numerically search for improved solutions is given by quantum
optimal control theory \cite{QOCT}, and will be the subject of
future investigations.

\section*{Acknowledgments}

This work was performed with the support of the EC project ACQP
(IST-2001-38863). M. A. Cirone and A. Negretti acknowledge partial
financial support from the ESF program QIT. M. A. Cirone also
acknowledges financial support from the EU-funded project QUEST
and the friendly hospitality at ECT* in Trento. A. Negretti acknowledges the financial support provided through the European Community's Human Potential Programme under contract HPRN-CT-2002-00304, [FASTNet]. A. Negretti thanks
S. Bettelli and C. Henkel for very stimulating discussions and the friendly hospitality at Institut f\"ur Physik in Potsdam.

\section*{APPENDIX: FIDELITY}

First we see how it is possible to calculate the minimum of the
expression (\ref{traxfid}). Let us define the function
\begin{equation}
\label{lagrange}
\mathcal{L}\left(x_{\gamma}\right)=\sum_{\alpha,\beta}x_{\beta}x_{\alpha}T_{\beta\alpha},
\end{equation}
where $x_{\gamma}=\left|c_{\gamma}\right|^2$ with the constraint
given by the set of zeros of the function
\begin{equation}
\mathcal{G}\left(x_{\gamma}\right)=\sum_{\alpha}x_{\alpha}-1.
\end{equation}
Thus we have to solve the linear system of equations
\begin{equation}
\label{eqvet}
\nabla\mathcal{L}-\lambda\nabla\mathcal{G}=0.
\end{equation}

that is,
\begin{equation}
M_{\alpha\beta}x_{\beta}=\lambda\qquad M_{\beta\alpha}=T_{\beta\alpha}+T_{\alpha\beta}.
\end{equation}

The minimum of $\mathcal{L}$ is then
\begin{eqnarray}
\label{flam2}
\mathcal{L}&=&\frac{1}{2}\left[\sum_{\beta\alpha}x_{\beta}x_{\alpha}T_{\beta\alpha}+\sum_{\alpha\beta}x_{\alpha}x_{\beta}T_{\alpha\beta}\right]=\frac{1}{2}\sum_{\alpha\beta}x_{\beta}\left[T_{\beta\alpha}+T_{\alpha\beta}\right]x_{\alpha}=\nonumber\\&=&\frac{1}{2}\sum_{\beta}x_{\beta}\sum_{\alpha}M_{\beta\alpha}x_{\alpha}=\frac{1}{2}\sum_{\beta}x_{\beta}\lambda=\frac{\lambda}{2}.
\end{eqnarray}

\end{document}